\documentclass[aps,prl,amsmath,amssymb,doubledspace]{revtex4}
\usepackage{amsmath}
\usepackage{graphicx}
\usepackage{rotating}
\usepackage{units}
\usepackage{color}
\usepackage{ulem}
\usepackage{nicefrac}
\usepackage{placeins}
\newcommand{\ket}[1]{\left|{#1}\right\rangle}
\newcommand{\bra}[1]{\left\langle{#1}\right|}

\begin{document}

\title{Precise ultra fast single qubit control using optimal control pulses}

\author{Jochen Scheuer$^{1}$, Xi Kong$^{1,2}$, Ressa S. Said$^{3}$, Jeson Chen$^{4}$, Andrea Kurz$^{1}$, Luca Marseglia$^{1}$, Jiangfeng Du$^{2}$, Philip R. Hemmer$^{4}$, Simone Montangero$^{3}$, Tommaso Calarco$^{3}$}
\author{Boris Naydenov$^{1}$}
\email{boris.naydenov@uni-ulm.de}
\author{Fedor Jelezko$^{1}$}
\affiliation{
$^1$Institut f\"ur Quantenoptik, Albert-Einstein-Allee 11, Universit\"at Ulm, 89069 Ulm, Germany\\
$^2$Department of Modern Physics, University of Science and Technology of China, Hefei, Anhui 230026, China\\
$^3$Institut f\"ur Quanteninformationsverarbeitung, Albert-Einstein-Allee 11, Universit\"at Ulm, 89069 Ulm, Germany\\
$^4$Electrical and Computer Engineering, Texas A\&M University, College Station, TX 77843, USA
}

\begin{abstract}
Ultra fast and accurate quantum operations are required in many modern scientific areas - for instance quantum information, quantum metrology and magnetometry. However the accuracy is limited if the Rabi frequency is comparable with the transition frequency due to the breakdown of the rotating wave approximation (RWA).  Here we report the experimental implementation of a method based on optimal control theory, which does not suffer these restrictions. We realized the most commonly used single qubit rotations - the Hadamard ($\pi/2$ pulse) and NOT ($\pi$ pulse) gates with fidelity  ($F^{\mathrm{exp}}_{\pi/2}=0.95\pm0.01$ and  $F^{\mathrm{exp}}_{\pi}=0.99\pm0.016$), in an excellent agreement with the theoretical predictions ($F^{\mathrm{theory}}_{\pi/2}=0.9545$ and $F^{\mathrm{theory}}_{\pi}=0.9986$). Moreover, we demonstrate magnetic resonance experiments both in the rotating and lab frames in this strong driving regime and we can deliberately "switch" between these two frames. Since our technique is general, it could find a wide application in magnetic resonance, quantum computing, quantum optics and broadband magnetometry.
\end{abstract}
\maketitle

\clearpage

The ability to manipulate spins very fast allows to increase the number of qubit operations before detrimental effects of decoherence take place, and it may further increase the bandwidth of spin based magnetometers. Optimal control theory \cite{Khaneja01,Khaneja2005} provides powerful tools to work in this regime by finding the optimal way to transform the system from the initial to the desired state with a high fidelity. Here we demonstrate precisely controlled ultra-fast (beyond RWA) single electron spin gates using specially designed microwave fields, without resorting to the standard RWA condition. To achieve this we employ optimal control namely the Chopped Random Basis~(CRAB) quantum optimization algorithm~\cite{Doria2011,Caneva2011}. It is used to numerically design and optimize our microwave control (see Supplementary information for more details). Our current algorithm uses a simple derivative-free direct search method to perform a multivariable function optimization~\cite{Caneva2011}, and hence offers enhanced computational flexibility such as parallel numerical calculations.

Here we implement this method using the electron spin associated with a single Nitrogen-Vacancy center (NV) in diamond as a test system. NV shows remarkable physical properties: optical spin initialization and readout of a single center at room temperature~\cite{Gruber97}, coherent spin control via microwaves (MW)~\cite{Jelezko2004a}, and a coherence time of several milliseconds~\cite{Balasubramanian2009}. This system is very promising as a nano-scale  ultra sensitive magnetometer~\cite{Taylor2008,Maze2008,Waldherr2012,Reinhard13} and solid state qubit~\cite{Neumann2008,Neumann2010}. Strongly driven dynamics of NV using conventional pulses has been also reported ~\cite{Fuchs2009}, where fast flipping of the NV's electron spin has been demonstrated.

\begin{figure}[htbp] 
\begin{center}
\includegraphics[scale=0.22]{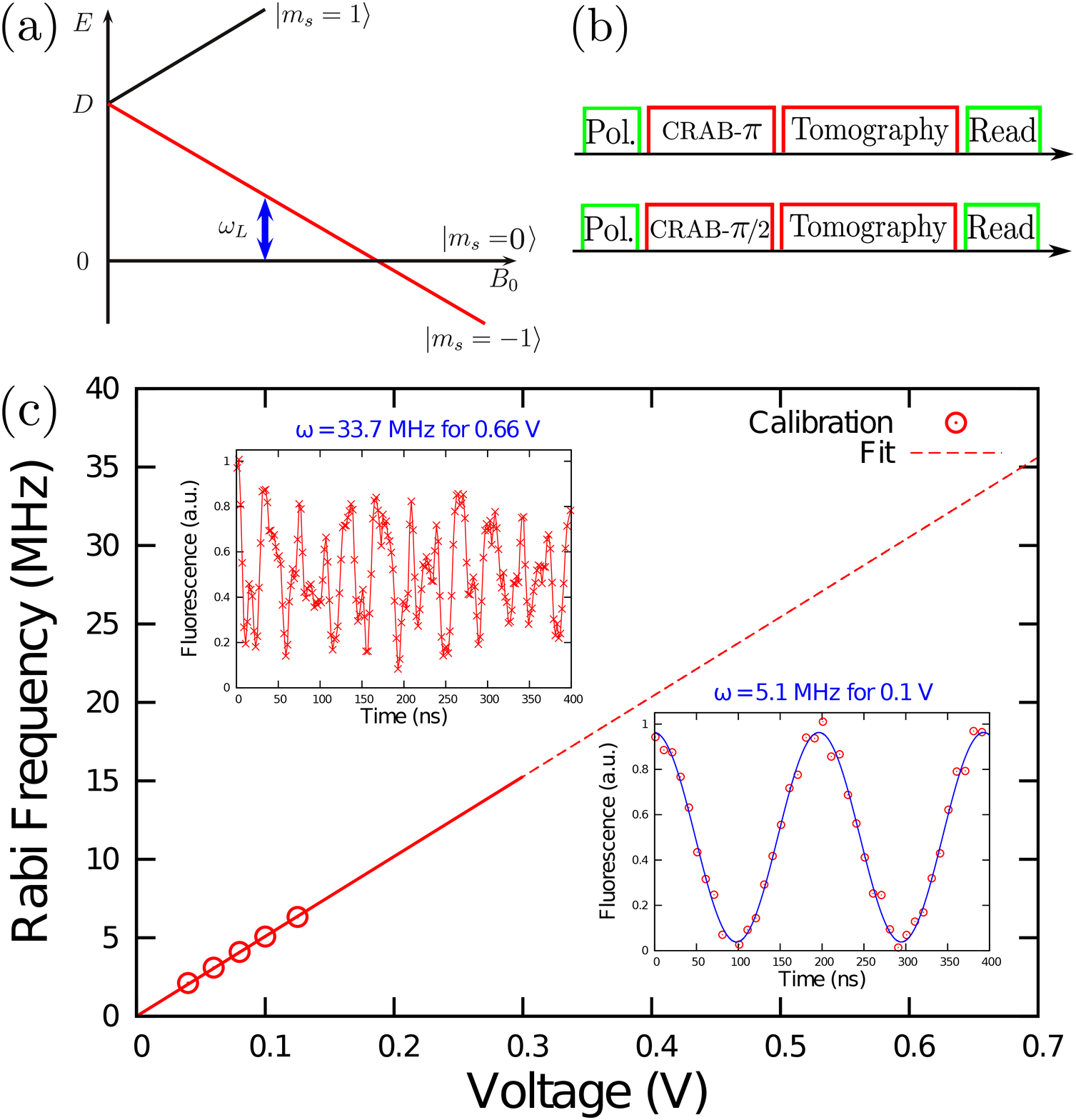}
\end{center}
\caption{(a) Energy of the $\ket{m_s}$ states of the NV center as a function of the applied static magnetic field $B_0$. $\omega_L=30$~MHz is the frequency of the transition we used in our experiments. (b) Schematic representation of the pulse sequences for the density matrix tomography. At the beginning we apply a laser pulse to polarize the NV in $\vert m_s=0\rangle$ and at the end again to read out the state of the electron spin. This is followed by the optimized $\pi$ pulse (CRAB-$\pi$, top) and $\pi/2$ (CRAB-$\pi/2$,bottom) pulses. The tomography is performed by applying $\pi/2$ pulses along the $x$ and $y$ axis of the rotating frame. (c) Rabi frequency as a function of the amplitude of the AWG signal. The circles denote the region where harmonic behavior is observed as shown in the lower right inset. The red line is a linear fit, its dashed part shows the region where the spin dynamics is anharmonic. The upper left inset shows a typical signal in this region. We chose a MW amplitude which corresponds to a Rabi frequency of $\Omega=30$~MHz.}
\label{PulseSequences}
\end{figure}

The NV consists of a substitutional nitrogen atom and an adjacent vacancy with a triplet ground state ($S=1$) and a strong optical transition, enabling the detection of single centers. Its fluorescence depends on the electron spin state, allowing one to perform coherent single spin control~\citep{Gruber97,Jelezko2004a}. The Hamiltonian of the NV's ground state in the presence of MW control~$\Gamma_x(t)$ can be written as:
\begin{eqnarray}
\hat{\mathcal{H}}/\left(2\pi\hbar\right) = D \hat{S_z^2} + \omega_z \hat{S}_z + \sqrt{2} \Gamma_x(t) \hat{S}_x,
\end{eqnarray}
where $D\approx 2.87$~GHz is the NV's electron zero-field splitting (ZFS), $\hat S_x$ and $\hat S_z$ are the $x$ and $z$ components of the electron spin operator and $\omega_z=g\mu_BB_0/\hbar$ is the Zeeman splitting due to a constant magnetic field~$B_0$ along the NV axis ($z$ axis) with $g$ the electron gyromagnetic ratio and $\mu_B$ the Bohr magneton~\cite{Wrachtrup2006b}. For the description of our experiments the Hamiltonian has to be written in the lab frame since the control amplitude is comparable to the Larmor frequency of the spin~\mbox{$\max \{|\Gamma_x(t)|\} \sim \omega_L$}~(where \mbox{$\omega_L = D- \omega_z$}), i.e. the counter-propagating term of the control can not be neglected~\cite{Fuchs2009}. To fulfill this condition beyond the RWA and to approximate a two-level spin system~of~\mbox{$|m_s=0\rangle$ and $|m_s=-1\rangle$}, we apply a magnetic field $B_0=1017.3$~G ($101.73$~mT) and set the working transition frequency $\omega_L=30$~MHz (Fig.~\ref{PulseSequences}a). $B_0$ is aligned along the NV crystal axis in order to suppress mixing between the $\ket{0}$ and $\ket{1}$ states.

\begin{figure}[htbp]
\centering
\includegraphics[scale=0.2]{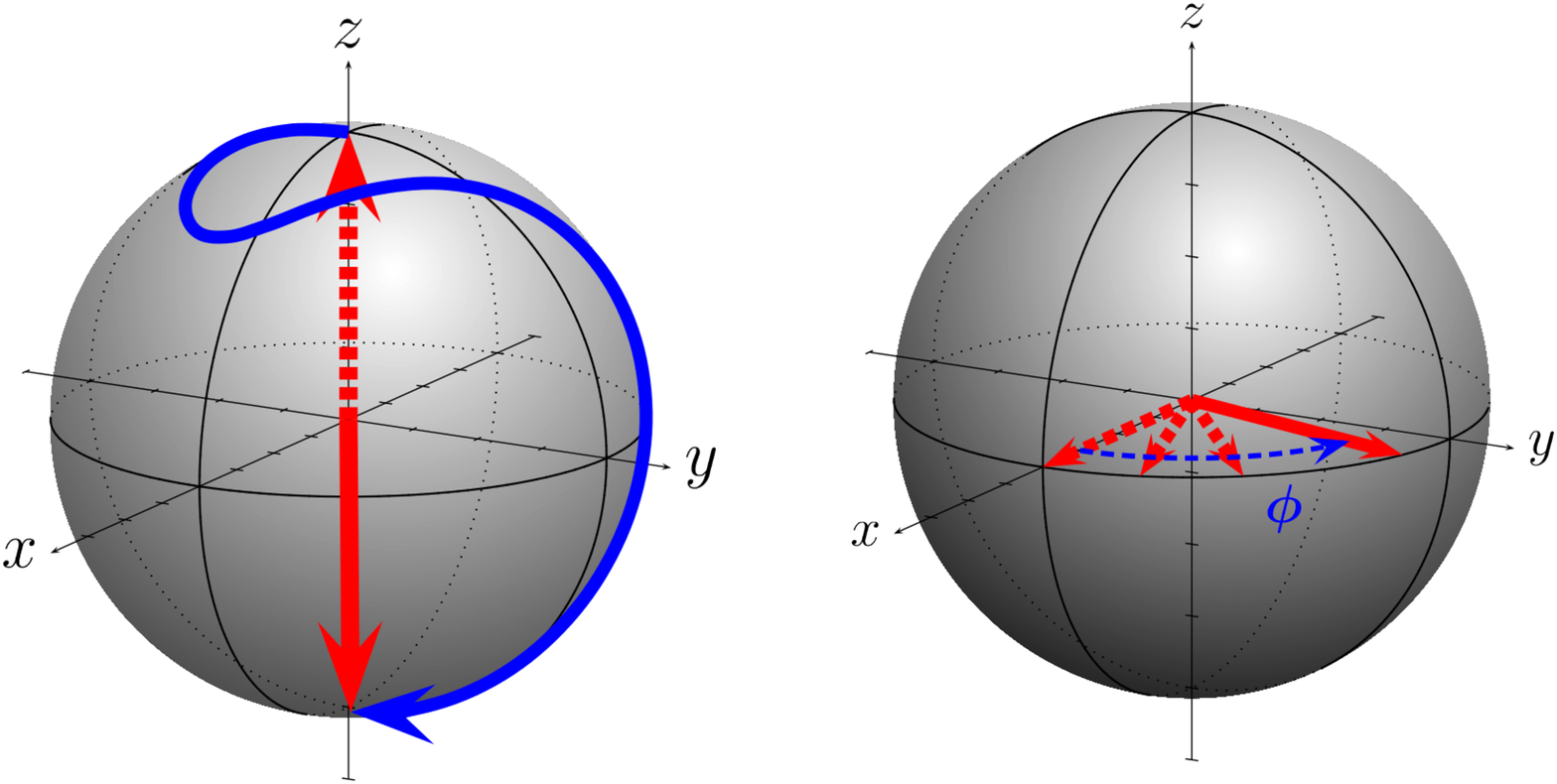}
\caption{(left) The trajectory of the spin magnetization (blue curve) during the application of the CRAB-$\pi$ pulse. The initial state is $m_s=0$ (red dashed arrow) and the target state is $m_s=-1$ (red solid arrow). (right) After the CRAB-$\pi/2$ the spin magnetization lays in the $xy$ plane of the lab frame, parallel to the $x$ axis. Then it rotates around the $z$ axis with an angular velocity $\omega_L$ (Larmor frequency), acquiring a phase $\phi=e^{-i\omega_L t}$.}
\label{BlochSpheres}
\end{figure}

We begin by measuring Rabi oscillations at different MW amplitudes and observe the spin dynamics (Fig.~\ref{PulseSequences}c). When the driving field $\Omega<\omega_L/2$ the system is in the RWA regime and a nice harmonic signal is obtained  (figure~\ref{PulseSequences}c, lower right inset). However, when $\Omega>\omega_L/2$ the signal becomes anharmonic as shown in Fig.~\ref{PulseSequences}c (upper left inset) and the precise control over the spin rotations is not trivial. Moreover, often the spin is not flipped within certain time as it can be seen from the upper left inset of Fig.~\ref{PulseSequences}c, where the normalized fluorescence does not go to zero and the  $\ket{m_s=-1}$ state is not reached.\\
To perform a desired transformation of the spin state $\ket{\psi(t)}$, which follows the Schr\"odinger equation, \mbox{$d|\psi(t)\rangle/dt = -i\hat{\mathcal{H}} |\psi(t)\rangle$} (assuming~$\hbar$=1), we optimally engineer the time dependence of the control $\Gamma_x(t)$, such that at the final time $T$ the state fidelity between the final spin state $|\psi(T)\rangle$ and the target $|\psi_T\rangle$ is maximized (see also the supplementary information). The fidelity $F$ is defined as~\cite{Nielsen2000}:
\begin{equation}
F = \sqrt{\langle\psi_{\mathrm{target}}\vert\rho_{\mathrm{CRAB}}\vert\psi_{\mathrm{target}}\rangle}
,\label{Fidelity}
\end{equation}
with $\psi_{\mathrm{target}}$ and $\rho_{\mathrm{CRAB}}$ being respectively the target state and the state expected after the CRAB pulse.\\
Here we used CRAB controls to implement the two most important single-qubit rotations: flipping the qubit (NOT-gate, $\pi$ pulse) and creating a superposition between the qubit states (Hadamard gate, $\pi/2$). The latter requires an additional rotation around the $z$ axis by $\pi/4$, which leads to phase change, irrelevant for a quantum algorithm \cite{MehringQC}. The experimental realization of these rotations is shown schematically in Fig.~\ref{PulseSequences}(b) while Fig.~\ref{BlochSpheres} shows the calculated trajectory of the spin during the CRAB-$\pi$ pulse.  
We set the pulse lengths $T_{\pi}=15.4$~ns and $T_{\pi/2}=7.7~$ns, shorter than  $T^{\mathrm{Rabi}}_{\pi}=1/2\Omega=16.67~$ns and $T^{\mathrm{Rabi}}_{\pi/2}=1/4\Omega=8.33~$, where $\Omega=30$~MHz is the extrapolated Rabi frequency, see Fig.~\ref{PulseSequences}. It is important to note that the latter cannot be used for qubit rotations due to the breakdown of the RWA (see Fig.~\ref{PulseSequences}). The shortest possible pulse length for the $\pi$ pulse is given by the Bang-Bang condition $T^{\mathrm{Bang}}_{\pi}= \pi/\sqrt{(\pi\omega_L)^2 + (2\pi\Omega)^2}=14.9$~ns~\cite{Boscain2006}. We are not able to reach this value due to the limited bandwidth and gain curve of the amplifier (see also Supplementary Information).

We performed a density matrix tomography in order to determine the quality of the optimized pulses. Both off-diagonal elements have been measured by applying a low power ($\Omega=8$ MHz) $\pi/2$ pulse along the $x$ and $y$ axis of the rotating frame, followed by a laser pulse for read out (see Fig.~\ref{PulseSequences}b). To measure the diagonal elements the  MW pulses have been omitted. After the CRAB-$\pi$ pulse the theoretically expected and the experimentally measured density matrices are:
\begin{eqnarray*}
\rho_{\mathrm{theory}}^{\pi}=\left(
\begin{array}{cc}
0 & 0 \\
0 & 1 \\
\end{array}
\right);
\rho_{\mathrm{exp}}^{\pi}=\left(
\begin{array}{cc}
0.01 & 0.16 - 0.15i \\
0.16+ 0.15 i & 0.99 \\
\end{array}
\right).
\end{eqnarray*}
After the CRAB-$\pi/2$ pulse we expect:
\begin{eqnarray*}
\rho_{\mathrm{theory}}^{\pi/2}=\left(
\begin{array}{cc}
0.5 & 0.5 \\
0.5 & 0.5 \\
\end{array}
\right)
\end{eqnarray*}
This is the state of the system expected directly after the MW pulse. However, due to limited time resolution of our apparatus, the measurement can be performed only after some dead time. We set this time to $t_{\mathrm{evol}}=100$~ns during which the spin rotates in the $xy$ plane in the lab frame and acquires a phase $\phi=e^{-i\omega_L t_{\mathrm{evol}}}$ (see Fig.~\ref{BlochSpheres}, right). The density matrix after $t_{\mathrm{evol}}$ is then:
\begin{eqnarray*}
\rho_{\mathrm{theory}}^{\pi/2}=\left(
\begin{array}{cc}
0.5 & 0.48 - 0.14i \\
0.48 + 0.14i & 0.5 \\
\end{array}
\right)
\end{eqnarray*}
From the tomography we obtain:
\begin{eqnarray*}
\rho_{\mathrm{exp}}^{\pi/2}=\left(
\begin{array}{cc}
0.57 & 0.42 - 0.02i \\
0.42 + 0.02i & 0.43 \\
\end{array}
\right)
\end{eqnarray*}
The expected fidelities of the CRAB pulses are $F_{\mathrm{theory}}^{\pi} = 0.9986$ and $F_{\mathrm{theory}}^{\pi/2} = 0.9545$, whilst from the experiment we obtain $F_{\mathrm{exp}}^{\pi} = (0.993\pm0.016)$ and $F_{\mathrm{exp}}^{\pi/2} = (0.947\pm0.007)$. We find an excellent agreement between the theoretical prediction and the experimental result. The theoretical values are lower than 1 due to the constrains set on the pulse duration. If the pulse length is increased, much higher numerical fidelities can be achieved \cite{Fortunato02}. The deviation from the prediction can be explained by distortion of the pulse shape, mainly due to the limited bandwidth of the MW amplifier and measurement error (see supplementary information).\\
The pulses we have developed in this study are important not only for quantum information processing, but also for most of the pulsed Nuclear Magnetic Resonance (NMR) and Electron Spin Resonance (ESR) experiments. Although they were not specifically designed as gates, only to transfer the spin from $\vert m_s =0\rangle$ to some desired state, they are also very robust and can be used for magnetic resonance as we show below. One of the most important NMR (and ESR) pulse sequences consists of a single $\pi/2$ pulse, where the spin magnetization is rotated from the $z$-axis to the $xy$ plane in the rotating frame. The spins then precess and can be detected by the NMR detector thus giving a Free Induction Decay (FID) signal. The Fourier transform of the latter is the spectrum of the sample~\cite{Slichter96, Schweiger2001}. This experiment has been already implemented using NV and can be applied for detecting DC magnetic fields~\cite{Maze12,Suter13,Nusran2012}. Here, we show that we can perform it both in the lab and in the rotating frame by using both CRAB and conventional rectangular pulses. The main difference here compared to previously reported studies, e.g. \cite{Nusran2012, Fuchs12, Hanson12}, is that we can perform these experiments beyond the RWA.
\begin{figure}[htbp]
\centering
\includegraphics[width=0.45\textwidth,keepaspectratio]{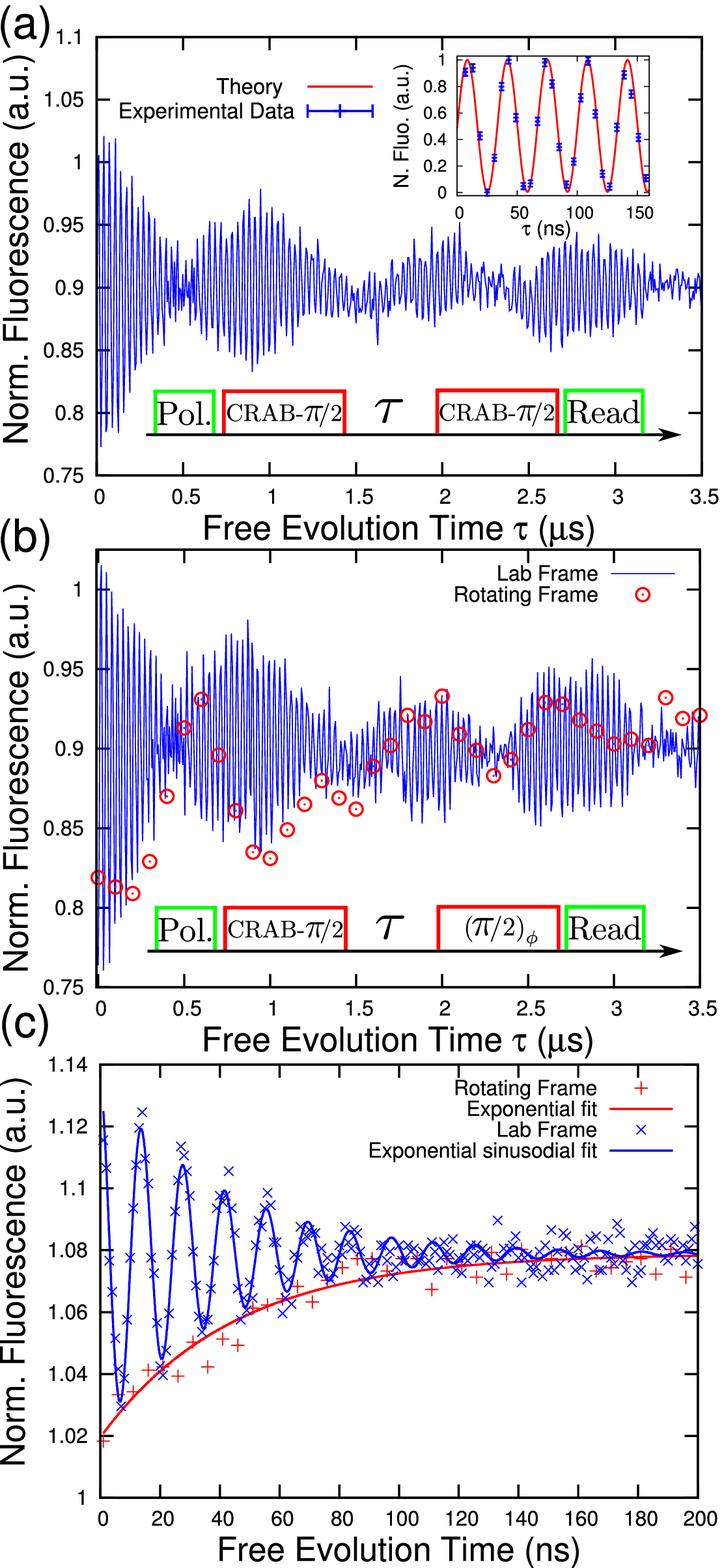}
\caption{Free Induction Decays. (a) FID measured by using two CRAB-$\pi/2$ pulses. The inset shows the first 160 ns of the signal (markers) and the calculated fidelity with respect to the $\vert 0\rangle$ state (red solid line, see also text)(b) A combination of a CRAB-$\pi/2$ pulse and a low power $\pi/2$ pulse, where he phase of the latter is fixed for all values of $\tau$ (blue curve) and adjusted as $\phi=e^{-i\omega_L t}$ (red markers), see also the main text. The oscillations with $\approx$ 2 MHz come from a weakly coupled $^{13}$C. (c) FID measured by using low power MW pulses (within RWA), where the phase of the second $\pi/2$ pulse is not in phase with the first pulse (blue markers). If both pulses are kept in phase then the experiment is performed in the rotating frame (red markers). The solid lines are fits to the data. The decay time here is shorter than in the above experiments because the sample is different.}
\label{FIDs}
\end{figure}
All sequences begin with a laser pulse for polarizing the NV (Fig.~\ref{FIDs}a). Then we apply a CRAB-$\pi/2$ pulse, which aligns the spin magnetization along the $x$ axis of the lab frame. After a free evolution time $\tau$ we apply another CRAB-$\pi/2$ pulse to rotate the spin back to the $z$ axis and we then read out the spin state. The signal oscillates with the Larmor Frequency $\omega_L$ (Fig.~\ref{FIDs}a, see also Fig.~\ref{BlochSpheres}, right). Here we measure the signal in the laboratory frame. Now, if we replace the second CRAB-$\pi/2$ pulse with a low power rectangular $\pi/2$ pulse (here $\omega_L=4193$~MHz and $\Omega=48$~MHz) and keep its phase fixed for all $\tau$ (equivalent to having the $B_1$ parallel to $y$), again the FID signal oscillates with $\omega_L$ (Fig.~\ref{FIDs}b, blue curve). However, if we set the phase  of the second pulse to $\phi=e^{-i\omega_L t}$, we can "follow" the magnetization in the equatorial plane of the Bloch sphere (Fig.~\ref{BlochSpheres}, right) and the measurement is performed in the rotating frame and the oscillations at $\omega_L$  disappear (Fig.~\ref{FIDs}b, red markers).\\
The same experiment can be performed using two rectangular pulses with low MW amplitude (within the RWA) where again the phase of the second pulse is kept constant. In this case the experiment is run in the laboratory frame and the FID again oscillates with the Larmor frequency (Fig.~\ref{FIDs} c, blue markers). If the phase of the MW is not the same for different $\tau$, then we obtain the typical FID in the rotating frame as shown in Fig.~\ref{FIDs}c (red markers). Thus we can on demand "switch" between the lab and rotating frames by using ultra-fast (beyond RWA) or conventional (lower power) pulses. Using this method in principle any oscillation of the spin signal can be "followed", while effectively canceling other unwanted frequencies. In the Supplementary Information we demonstrate this technique when the excitation is not in resonance.\\
\begin{figure}[htbp]
\begin{center}
\includegraphics[scale=0.75,keepaspectratio]{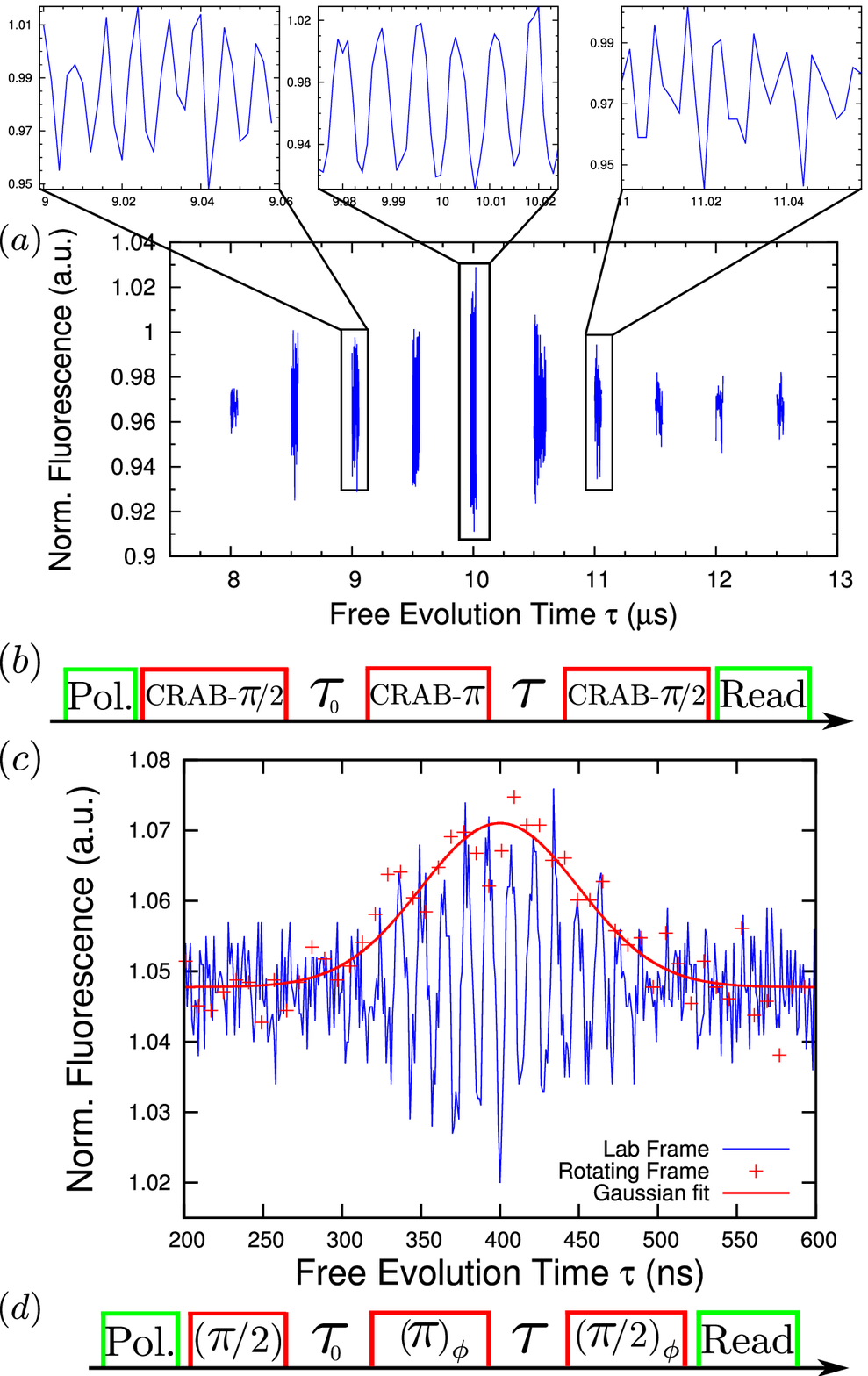}
\end{center}
\caption{ (a) Hahn echo measured with CRAB-pulses. This signal oscillates with the Larmor frequency (in this case $\omega_L=120$~MHz and the extrapolated Rabi frequency $\Omega=30$~MHz), as shown in the three insets. (b) CRAB-Hahn echo pulse sequence. (c) Hahn echo signal obtained with rectangular pulses in RWA when the first MW pulse is not kept in phase with the last two pulses (blue curve) The signal again oscillates with (in this case) $\omega_L=71.3$~MHz. The same experiment, here all pulses are kept in phase (red markers). The solid red line is a Gaussian fit to the data. (d) Hahn echo pulse sequence with rectangular pulses where the phase $\phi$ of the pulses is properly adjusted.}
\label{HahnEcho}
\end{figure}
Another important sequence is the Hahn echo~\cite{Hahn50}, which has found wide application in NMR and ESR. It is the basis of all dynamical decoupling techniques since all static inhomogeneous shifts (and fluctuations on the time scale of the coherence time $T_2$) are effectively canceled out~\cite{Slichter96}. It has been implemented with NV~\cite{Jelezko2004a} and also for NV based AC magnetometry~\cite{Maze2008,Balasubramanian2009}. The Hahn echo pulse sequence is depicted in Fig.~\ref{HahnEcho}b. After a CRAB-$\pi/2$ pulse and a free evolution time $\tau_0$ a CRAB-$\pi$ pulse is applied. After a time $\tau$ the spin state is rotated to the $z$ axis by a CRAB-$\pi/2$ pulse and then it is read out by a laser pulse. The spin signal oscillates with the Larmor frequency of the NV transition (in this case $\omega_L=120$~MHz) as shown in Fig.~\ref{HahnEcho}a since the experiment is performed in the lab frame. If we use rectangular pulses in the rotating frame we obtain the envelope of the echo without the oscillations at $\omega_L$ (Fig.~\ref{HahnEcho}c, red markers). However, we can again "switch" to the lab frame by adjusting the phase $\phi$ of the last MW pulse (Fig.~\ref{HahnEcho}c, blue curve). ESR experiments in the lab frame with low power MW pulses have been already reported~\cite{Schweiger04, Eaton11}, but the "switching" between the two frames in the strong driving regime has not been yet demonstrated. Our method presented here allows us to obtain full control over the phase of a spin system and can be implemented with any other pulse sequences.\\
In summary, we have developed a novel technique based on optimal control for precise spin qubit rotations in the ultra fast driving regime where standard pulses are not applicable. We design our qubit gates by using the quantum optimization algorithm CRAB and find an excellent agreement with the experimental implementation. Moreover, we show that the pulses developed here can be used for basic magnetic resonance experiments also in the case when the rotating frame approximation breaks down. Additionally we demonstrate on demand "switching" between the rotating and the lab frame, using both CRAB and conventional (low power) pulses. This provides a precise control over the spin evolution and can be easily transferred to any other two level system, e.g. trapped atoms, trapped ions or superconducting qubits.  Our results can find wide application in quantum computation and broadband magnetometry.\\
\FloatBarrier
\begin{acknowledgements}

We thank Florian Dolde for the experimental support. This work is supported by EU STREP Project DIAMANT, SIQS, DIADEMS, DFG (SPP 1601/1, SFB TR21) and BMBF (QUOREP). Numerical optimization were performed on the bwGRiD (http://www.bw-grid.de). 
\end{acknowledgements}


\newpage
\begin{center}
{\huge \bf{Precise ultra fast single qubit control using designed quantum gates\\}}
\vspace{2 cm}
Jochen Scheuer$^{1}$, Xi Kong$^{1,2}$, Ressa S. Said$^{3}$, Jeson Chen$^{4}$, Andrea Kurz$^{1}$, Luca Marseglia$^{1}$, Jiangfeng Du$^{2}$, Philip R. Hemmer$^{4}$, Simone Montangero$^{3}$, Tommaso Calarco$^{3}$, Boris Naydenov$^{1}$ and Fedor Jelezko$^{1}$
\vspace{2 cm}

$^1$Institut f\"ur Quantenoptik, Albert-Einstein-Allee 11, Universit\"at Ulm, 89069 Ulm, Germany\\
$^2$Department of Modern Physics, University of Science and Technology of China, Hefei, Anhui 230026, China\\
$^3$Institut f\"ur Quanteninformationsverarbeitung, Albert-Einstein-Allee 11, Universit\"at Ulm, 89069 Ulm, Germany\\
$^4$Electrical and Computer Engineering, Texas A\&M University, College Station, TX 77843, USA
\end{center}
\newpage
 \tableofcontents
\section{CRAB  Microwave Control}

In this section we shortly review the theoretical background of the chopped random basis (CRAB) method in the context of our current work, and furthermore describe the optimization procedures and the numerical results.  

We recall the ground state Hamiltonian of the single electron NV spin in the presence of a single MW control~$\Gamma_x(t)$, as discussed in the manuscript, 
\begin{eqnarray}
\hat{\mathcal{H}}/\left(2\pi\hbar\right) = D \hat{S_z^2} + \omega_z \hat{S}_z + \sqrt{2} \Gamma_x(t) \hat{S}_x.
\end{eqnarray}

The CRAB method designs the control $\Gamma_x(t)$ by correcting an initial guess $\Gamma_0(t)$ with an optimized continuous function~$g(t)$, following~\mbox{$\Gamma_x(t)=\Gamma_0(t)\times g(t)$}~\mbox{\cite{Doria2011,Caneva2011}}. We use here a simple constant initial guess~\mbox{$\Gamma_0(t)=g_0$}. 
Following~\cite{Caneva2011}, the correcting function~$g(t)$ is expanded into a Fourier-like basis function,
\begin{eqnarray}
g(t) = \frac{1}{2N\lambda(t)} \sum_{n=1}^{N} \{ a_n \sin \left(\omega_n t \right) + b_n \cos \left( \omega_n t \right) \}, 
\end{eqnarray}
where $N$ denotes a number of basis expansion having $N$-randomized discrete frequencies. It is noteworthy to state that the range of frequencies~$(\omega_1,\omega_N)$ directly corresponds to the real bandwidth of the apparatus. Therefore, we can pre-set in advance the frequency range for numerical optimization to meet the experimental limitations, e.g. the amplifier's or arbitrary waveform generator's working bandwidth. 
The additional function $\lambda(t)$, is used to impose the control boundary such that~\mbox{$\Gamma_x(t) = 0$} at the initial time~$t = 0$, and the final time~$T$. Here, we choose the bounding function~\mbox{$\lambda(t) = h^p / (h^p - (t-h)^p)$}, where~$h=T/2$.  Using this function we can also qualitatively vary the rising and falling times of the microwave control by adjusting the even-numbered parameter $p$.

Our optimization objective is to find the set of CRAB parameters~\mbox{\{$\vec{a}_n,\vec{b}_n,\vec{\omega}_n$\}}, which minimizes the figure of merit,~\mbox{$\mathcal{F}= (1 - f) + c_f \max \{|\Gamma_x(t)|\}$}, where the quantity $f=|\langle \psi(T) | \psi_T \rangle|^2$ is the fidelity of the final state $|\psi(T)\rangle$, against the desired state $|\psi_T\rangle$. A dimensionless parameter~$c_f$ is incorporated in the figure of merit to limit the control amplitude during optimization. We employ the direct search simplex (Nelder-Mead) algorithm to find the optimal CRAB parameters. 

The numerical optimization is initiated by setting some parameters obtained from the experimental preparations and apparatus calibrations: the measured Larmor transition $\omega_L$, the maximum control amplitude~\mbox{$\max\{|\Gamma_x(t)|\}=\Omega$}, and the CRAB frequency range. The control time is fixed (the same as the desired rotation time) to be faster than the extrapolated rotation time if the RWA would be valid, e.g. for the spin~\mbox{$\pi-$rotation} we have~\mbox{$T<$\nicefrac{1}{2}$\Omega^{-1}$}, where~$\Omega$ is the extrapolated Rabi frequency (see figure 1c).  However, the \mbox{$\pi-$rotation} time in our case is limited by the minimum time of the theoretically proposed optimal {\it bang-bang} control,~\mbox{$T^{\mathrm{Bang}}_{\pi} =  \pi/\sqrt{(\pi\omega_L)^2 + (2\pi\Omega)^2}$}, where the control $\Gamma_x(t)$ takes only a constant value of $\pm \Omega$ for a certain time interval \cite{Boscain2006}.

To obtain the optimized pulse for one target rotation, we:
\begin{enumerate}
\item
Perform the parallel simplex search algorithm with an $S$ number of random initial values of the CRAB parameters for $j$ small positive real numbers of~$\{c_f^j\}$, and $k$ positive small integers $\{N^k\}$, typically $\{c_f^j\}\in (0.01,0.5)$ and $\{N^k\}\in (3,7)$.
\item
Obtain from Step 1 $(S\times j \times k)$ sets of CRAB parameters, than construct $(S\times j \times k)$ numbers of control pulses $\{\Gamma_x(t)\}$. 
\item
Investigate the numerical values of $\mathcal{F}$ and $\max \{|\Gamma_x(t)|\}$ for each pulse, and pick the best one out of $(S\times j \times k)$ pulses which satisfies $\mathcal{F} \leq \kappa_f$ and $\max \{|\Gamma_x(t)|\} \leq \kappa_\Gamma$. The preset quantities $\kappa_f$ and $\kappa_\Gamma$ are the numerical infidelity and the maximum control amplitude, respectively.
If the best pulse can not be obtained, return to Step 1 with different values of $\{c_f^j\}$ and increase $\{N^k\})$
%
\end{enumerate}

The actual numerical calculations were carried out on the bwGRiD cluster where we utilized its multi-nodes multi-cores (8 cores per node) computational features to run the parallel Nelder-Mead searches that corresponds to various experimental parameters and random initial values. To find a single parallel sample in one-core, i. e. one set of CRAB optimized parameters, the typical computational time required to meet the experimentally acceptable fidelity is approximately less than 30 minutes. This allows one to perform a single optimization run in just a decent commercial personal computer. Hence, it is feasible in the future to apply our numerical CRAB optimization in standard close-loop control system involving directly the control apparatuses. For both cases of $\pi-$rotation and $\nicefrac{\pi}{2}-$rotation we set the parameters as the following: $N=5$, $S=30$, $\omega_L=g_0=\max\{|\Gamma_x(t)|\}=30$~MHz, and $\omega_n\in (10,100)$ MHz. We present the best obtained CRAB parameters for each rotation in Table \ref{TableData}, while the corresponding MW pulses are shown in figure \ref{CRAP_PulseShape}.

\begin{table}[H]
\begin{center}
  \begin{tabular}{| c | c | c || c | c | c |}
       \hline
       \multicolumn{3}{|c||}{$\pi$-rotation} & \multicolumn{3}{c|}{$\nicefrac{\pi}{2}$-rotation} \\ \hline    
       \multicolumn{3}{|c||}{$T$ = 15.4071 ns, $p$ = 60, $c_f$ = 0.35.}  & \multicolumn{3}{c|}{$T$ = 7.7036 ns, $p$ = 38, $c_f$ = 0.23.} \\      
       \hline
  $a_n$ & $b_n$ & $\omega_n$ (GHz) & $a_n$ & $b_n$ & $\omega_n$ (GHz)\\ \hline     
  $a_1=$ -5.4865 & $b_1=$ 0.2812   & $\omega_1=$ 0.0201 & $a_1=$2.1123 & $b_1=$ 9.6205   & $\omega_1=$ 0.0149\\ 
  $a_2=$ 2.4803  & $b_2=$ 1.8823   & $\omega_2=$ 0.0415 & $a_2=$ -5.5973 & $b_2=$ -28.7365   & $\omega_2=$ 0.0401\\   
  $a_3=$ -0.5404 & $b_3=$ 5.8533   & $\omega_3=$ 0.0513 & $a_3=$ -9.7577 & $b_3=$ -3.9425   & $\omega_3=$ 0.0464\\ 
  $a_4=$ 1.5659  & $b_4=$ -2.2123  & $\omega_4=$ 0.0687 & $a_4=$26.3464 & $b_4=$5.4267   & $\omega_4=$ 0.0664\\ 
  $a_5=$ 1.4673  & $b_5=$  3.6469  & $\omega_5=$ 0.0892 & $a_5=$ -10.4212 & $b_5=$ 7.2445  & $\omega_5=$ 0.0909\\ \hline
  \end{tabular}
  \end{center}
\caption{The optimal CRAB parameters obtained via the Nelder-Mead simplex algorithm for $\pi$- and $\nicefrac{\pi}{2}$-rotations. The corresponding MW-CRAB pulses are shown in Fig. \ref{CRAP_PulseShape}.}
\label{TableData}
\end{table}
\section{Experimental Setup and Pulse Shapes}
\begin{figure}[H]
\centering
 \includegraphics[width=1\textwidth]{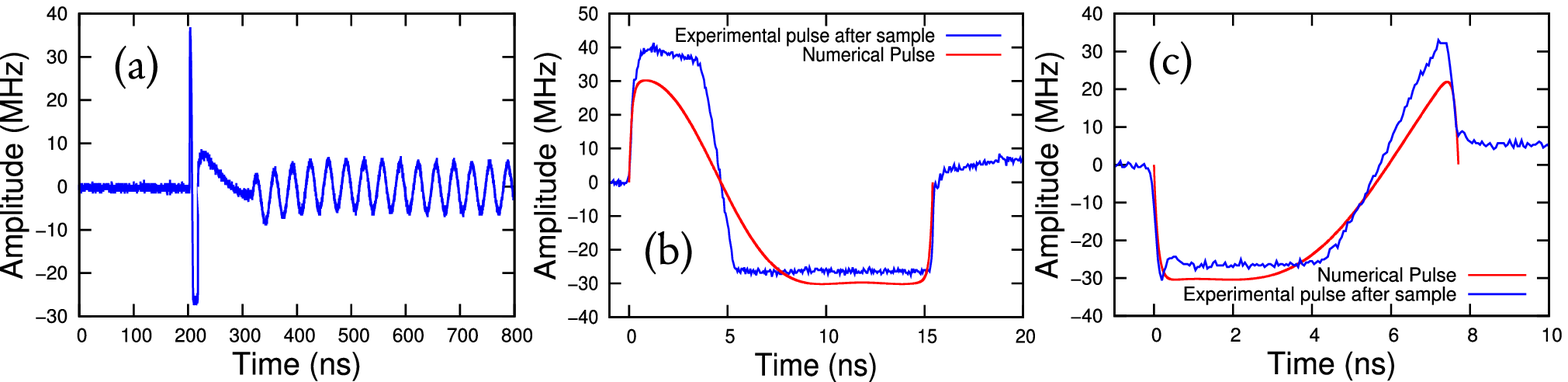}
\caption{Pulse Shapes. (a) Oscilloscope measurement of the signal after the diamond with conventional, sinusoidal microwave after 100$\,$ns delay to measure Rabi oscillations for the state tomography. (b) CRAB-$\pi$ pulse (blue) in comparison to numerical pulse (red) (c) CRAB-$\pi$/2 pulse (blue) in comparison to numerical pulse (red).}
\label{CRAP_PulseShape}
\end{figure}

The pulses optimized by the CRAB algorithm contain a superposition of ten sinusoidal functions. Experimentally these pulses are synthesized directly by an Arbitrary Waveform Generator (Tektronix, AWG7122C) with a sampling rate of 24$\,$GS/s and than sent to an amplifier (Mini-Circuits, ZHL-42W-SMA).
The pulse shapes measured with an oscilloscope (Tektronix, TDS6804B) are displayed in Fig. \ref{CRAP_PulseShape}.
Optical measurements were obtained via a self-made confocal microscope, the AWG triggered both the acousto-optic modulator for laser pulse control and the photon count card (FastComtec: P7887).

\section{State tomography}
\begin{figure}[H]
\centering
 \includegraphics[width=1\textwidth]{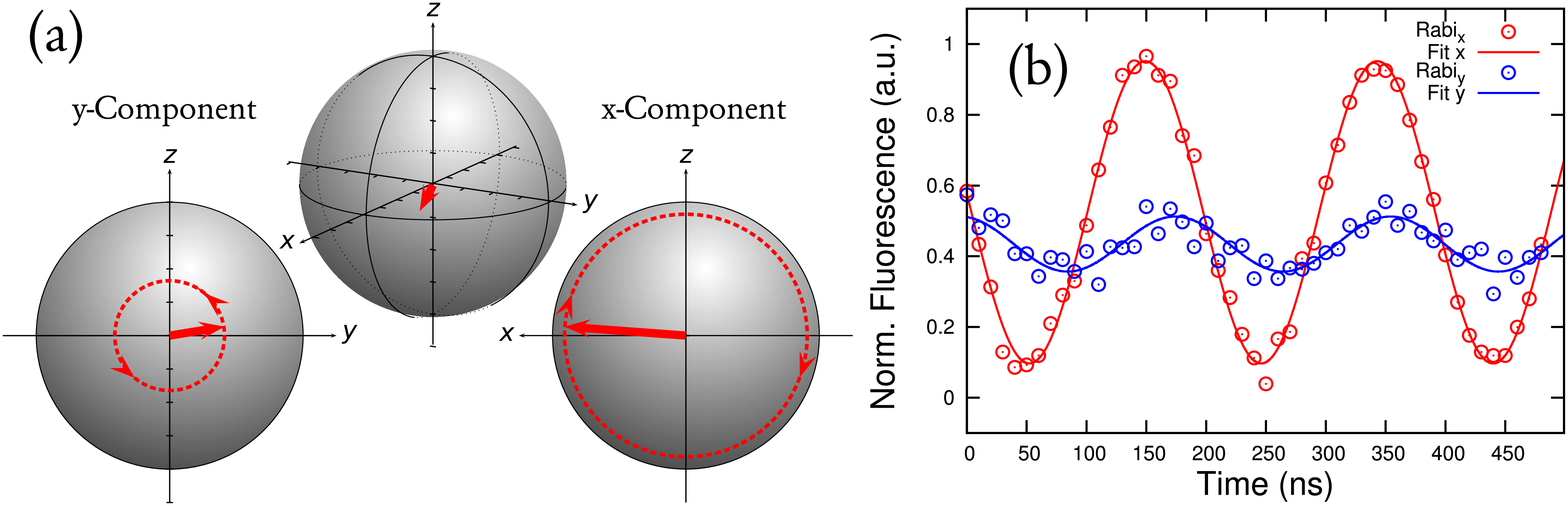}
\caption{State tomography (a) The state is characterized by performing Rabi oscillations by rotating the spin around the $x$- and $y$-axis. The $z$-component can be observed by measurement without application of a microwave. With the amplitude and the $z$-component the $x$- and $y$-component can be calculated by using the Pythagorean theorem. (b) Measurement data of the state tomography after the CRAB-$\pi/2$-pulse.}
\label{tomography}
\end{figure}

In order to determine the fidelity of the final state in respect to the target state we performed a state tomography. For this purpose we evaluated all three components of the Bloch vector.The $z$-component is measured directly from the fluorescence level.The $x$- and $y$-components have to be projected into the measurable $z$-component. This is done by applying a microwave with different phases and measure Rabi oscillations or in other words: rotate the Bloch vector around the $x$-axis to determine the $y$-component and vice versa, as shown in Fig. \ref{tomography}.

The density matrix of a single qubit can be written as follows:
\begin{align}
\rho = \frac{1}{2}\left( \begin{matrix} 1+z&x+i y\\ x-iy& 1-z \end{matrix} \right)\,.
\end{align}
In this definition, $x,y$ and $z$ have values between -1 and 1. The first point of the measurements $Rabi_x(1)$ and $Rabi_y(1)$, where no microwave was applied, can be understood as the $z$-component. The three components of the Bloch vector are calculated in the following way: 
\begin{align}
 z=2 \cdot \left< Rabi_x(1) + Rabi_y(1)\right> -1=Rabi_x(1) &+ Rabi_y(1) -1 \,,\\
  y=\sqrt{(2 Amp_y)^2-z^2}\,,&\label{Blochy}\\
 x=\sqrt{(2 Amp_x)^2-z^2}\,,&\label{Blochx}
 \end{align}
 
where $x,y$ and $z$ are the three components of the Bloch vector and $Amp_{x,y}$ are the amplitudes of the Rabi oscillations rotating around the $y$,$x$-axis respectively. In formula \ref{Blochy} and \ref{Blochx}, values for $x$ and $y$ were set to 0 if the uncertainty was greater than the actual value. Another possibility to obtain the $x$- and $y$-component is to calculate $Rabi_x(\pi/2)$ and $Rabi_y(\pi/2)$ respectively.

To normalize the data we performed an additional, bare Rabi measurement. The normalization is done by $Rabi_x(1)=(rawdata(1)-(y_0-A))/2A$, where $A$ is the amplitude and $y_0$ is the offset of the normalization measurement.

In order to calculate the fidelity between the experimental state $\rho$ and the target state $\ket{\Psi}$ the definition for pure states is used: $F=\sqrt{ \bra{\Psi}\rho \ket{\Psi}}$ which in this case is equivalent to the general definition $F=tr\sqrt{\sqrt{\sigma}\rho\sqrt{\sigma}}$~\cite{nielsenchuang}. 

Due to experimental limitations we had to wait for 100$\,$ns between the CRAB-pulse and the Rabi measurement, hence the target state after the time evolution on the $x,y$ plane becomes

\begin{align}
	\ket{\Psi(t)}=\text{e}^{-\frac{i}{2} \sigma_z \omega_L t} \ket{\Psi(0)}\,,
\end{align}

with the Pauli matrix $\sigma_z$ and the Larmor frequency $\omega_L$.
For the error calculation of the fidelity the noise of the Poisson distributed photon collection and fitting errors were taken into account. The error was determined by using the general law of error propagation~\cite{granicher1996messung}:
\begin{align}
  \Delta F = \sum \limits_{k=1}^{N} \left(\frac{\partial F}{\partial f_k} \right)^2 \cdot var(f_k) + 2 \cdot \sum \limits^{N-1}_{l=1} \sum \limits_{m=l+1}^{N} \left( \frac{\partial F}{\partial f_l} \right) \left(\frac{\partial F}{\partial f_m}  \right) \cdot cov(f_l,f_m)\,.
\end{align}

\section{Magnetic Resonance in an arbitrary frame - off resonant excitation}

\begin{figure}
\begin{tabular}{cc}
\includegraphics[width=0.49\textwidth]{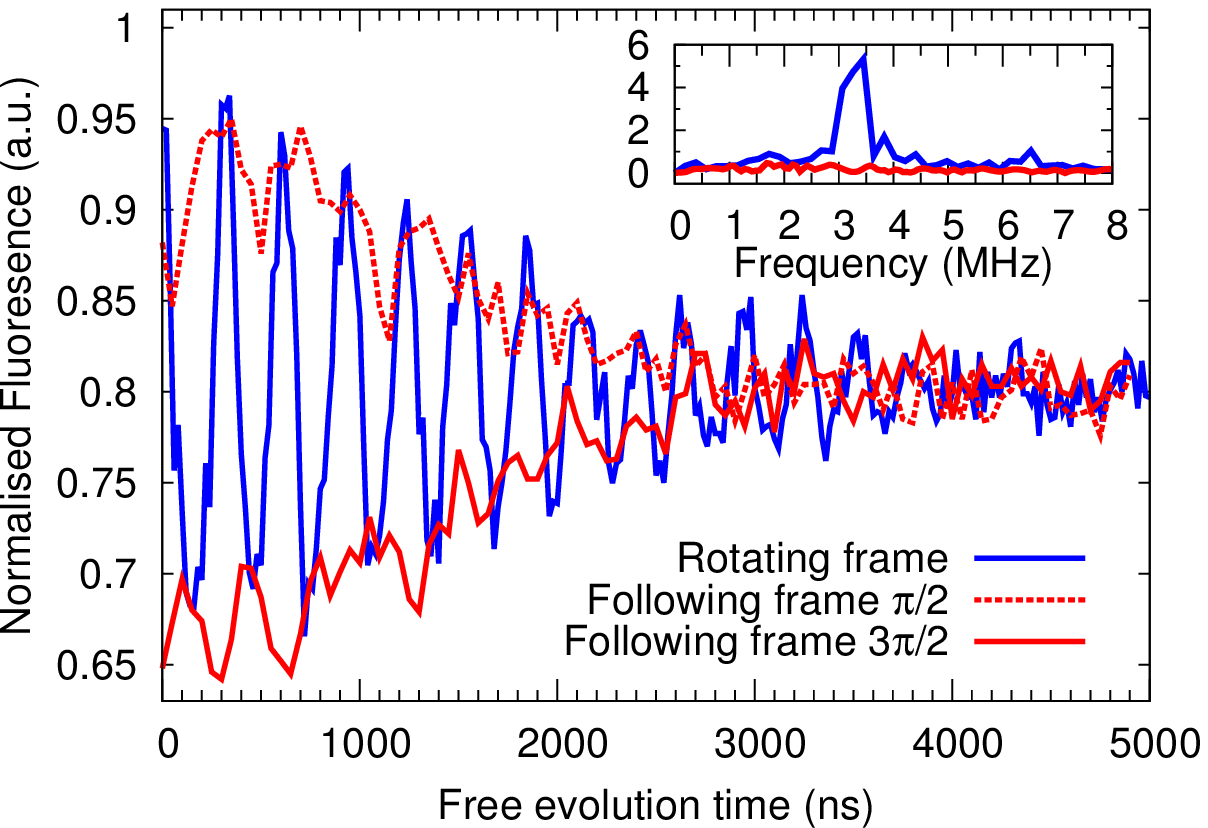}&
\includegraphics[width=0.49\textwidth]{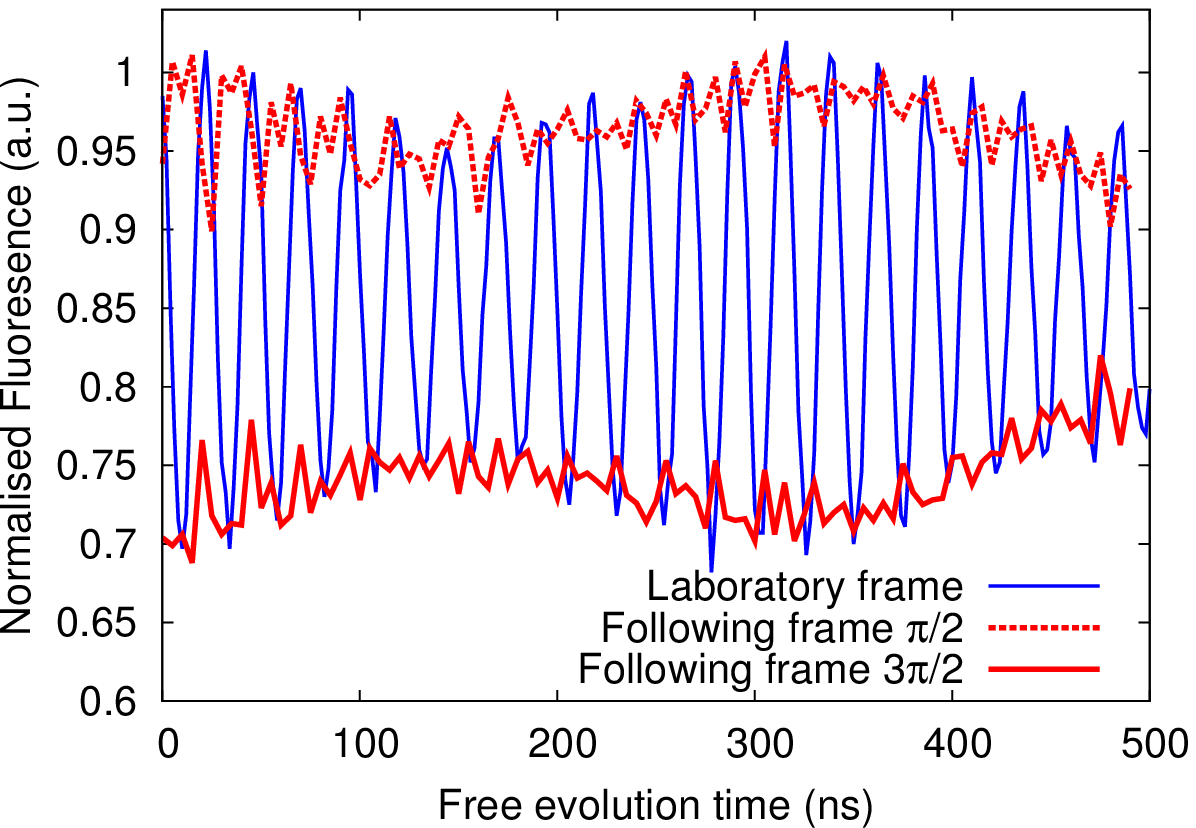}
\end{tabular}
\caption{(left) FID performed in the rotating (blue) and following (red) frames. The inset shows the Fourier transform of both signal. An oscillation with the off-resonance shift $\Delta\omega=3$~MHz is observed only when the experiment is performed in the rotating frame. (right) FID in lab and following frames. In the lab frame we observe a fast oscillation of the Larmor frequency and a slow beating due to the frequency shift $\Delta\omega$ (blue curve). If we compensate both Larmor and the $\omega$ (red data) we obtain the envelope of the lab frame measurement. For the solid (dashed) red lines the rotation angle of the second MW pulse was set to $\pi/2$ ($3\pi/2$).}
\label{Fig_Rot_Follow}
\end{figure}

Here we demonstrate that we are able to "switch" between different reference frames in the case of off-resonance excitation with an offset:
\begin{equation}
\Delta\omega = \omega-\omega_L
\end{equation}
where $\omega$ is the MW frequency and $\omega_L=40.8$~MHz is the Larmor frequency. In these experiments we set $\Delta\omega=3$~MHz. Now there are three coordinate frames, in all of them the $z$ axis is fixed in space and parallel to the constant magnetic field $B_0$. We distinguish between - laboratory frame (the $x$ and $y$ axis are fixed in space), rotating frame ($xy$ plane rotates with the Larmor frequency) and following frame ($xy$ plane rotates with the MW frequency $\omega=\Delta\omega+\omega_L$). We can measure FID in any of these frames by properly adjusting the phase of the second MW pulse (see also Fig. 3 in the main text). FID measured in the rotating frame at an off-resonant offset of $\Delta\omega=3$~MHz is shown in Fig.~\ref{Fig_Rot_Follow} (left, blue curve), where the Fourier transform of the signal confirms the expected frequency of 3 MHz. If the phase of the second MW $\pi/2$ pulse is set as $\phi=\Delta\omega\tau$, we are detecting the FID in the following frame, where no oscillation is observed, see Fig.~\ref{Fig_Rot_Follow} (left, red curves). By keeping the phase of both MW pulses fixed for values of $\tau$, we detect in the laboratory frame (Fig.~\ref{Fig_Rot_Follow}, right, blue curve.)


\begin{thebibliography}{29}
\expandafter\ifx\csname natexlab\endcsname\relax\def\natexlab#1{#1}\fi
\expandafter\ifx\csname bibnamefont\endcsname\relax
  \def\bibnamefont#1{#1}\fi
\expandafter\ifx\csname bibfnamefont\endcsname\relax
  \def\bibfnamefont#1{#1}\fi
\expandafter\ifx\csname citenamefont\endcsname\relax
  \def\citenamefont#1{#1}\fi
\expandafter\ifx\csname url\endcsname\relax
  \def\url#1{\texttt{#1}}\fi
\expandafter\ifx\csname urlprefix\endcsname\relax\def\urlprefix{URL }\fi
\providecommand{\bibinfo}[2]{#2}
\providecommand{\eprint}[2][]{\url{#2}}

\bibitem[{\citenamefont{Khaneja et~al.}(2001)\citenamefont{Khaneja, Brockett,
  and Glaser}}]{Khaneja01}
\bibinfo{author}{\bibfnamefont{N.}~\bibnamefont{Khaneja}},
  \bibinfo{author}{\bibfnamefont{R.}~\bibnamefont{Brockett}}, \bibnamefont{and}
  \bibinfo{author}{\bibfnamefont{S.~J.} \bibnamefont{Glaser}},
  \bibinfo{journal}{Phys. Rev. A} \textbf{\bibinfo{volume}{63}},
  \bibinfo{pages}{032308} (\bibinfo{year}{2001}).

\bibitem[{\citenamefont{Khaneja et~al.}(2005)\citenamefont{Khaneja, Reiss,
  Kehlet, Schulte-Herbr\"uggen, and Glaser}}]{Khaneja2005}
\bibinfo{author}{\bibfnamefont{N.}~\bibnamefont{Khaneja}},
  \bibinfo{author}{\bibfnamefont{T.}~\bibnamefont{Reiss}},
  \bibinfo{author}{\bibfnamefont{C.}~\bibnamefont{Kehlet}},
  \bibinfo{author}{\bibfnamefont{T.}~\bibnamefont{Schulte-Herbr\"uggen}},
  \bibnamefont{and} \bibinfo{author}{\bibfnamefont{S.~J.}
  \bibnamefont{Glaser}}, \bibinfo{journal}{J. Magn. Reson.}
  \textbf{\bibinfo{volume}{172}}, \bibinfo{pages}{296} (\bibinfo{year}{2005}).

\bibitem[{\citenamefont{Doria et~al.}(2011)\citenamefont{Doria, Calarco, and
  Montangero}}]{Doria2011}
\bibinfo{author}{\bibfnamefont{P.}~\bibnamefont{Doria}},
  \bibinfo{author}{\bibfnamefont{T.}~\bibnamefont{Calarco}}, \bibnamefont{and}
  \bibinfo{author}{\bibfnamefont{S.}~\bibnamefont{Montangero}},
  \bibinfo{journal}{Phys. Rev. Lett.} \textbf{\bibinfo{volume}{106}},
  \bibinfo{pages}{190501} (\bibinfo{year}{2011}).

\bibitem[{\citenamefont{Caneva et~al.}(2011)\citenamefont{Caneva, Calarco, and
  Montangero}}]{Caneva2011}
\bibinfo{author}{\bibfnamefont{T.}~\bibnamefont{Caneva}},
  \bibinfo{author}{\bibfnamefont{T.}~\bibnamefont{Calarco}}, \bibnamefont{and}
  \bibinfo{author}{\bibfnamefont{S.}~\bibnamefont{Montangero}},
  \bibinfo{journal}{Phys. Rev. A} \textbf{\bibinfo{volume}{84}},
  \bibinfo{pages}{022326} (\bibinfo{year}{2011}).

\bibitem[{\citenamefont{Gruber et~al.}(1997)\citenamefont{Gruber, Drabenstedt,
  Tietz, Fleury, Wrachtrup, and von Borczyskowski}}]{Gruber97}
\bibinfo{author}{\bibfnamefont{A.}~\bibnamefont{Gruber}},
  \bibinfo{author}{\bibfnamefont{A.}~\bibnamefont{Drabenstedt}},
  \bibinfo{author}{\bibfnamefont{C.}~\bibnamefont{Tietz}},
  \bibinfo{author}{\bibfnamefont{L.}~\bibnamefont{Fleury}},
  \bibinfo{author}{\bibfnamefont{J.}~\bibnamefont{Wrachtrup}},
  \bibnamefont{and} \bibinfo{author}{\bibfnamefont{C.}~\bibnamefont{von
  Borczyskowski}}, \bibinfo{journal}{Science} \textbf{\bibinfo{volume}{276}},
  \bibinfo{pages}{2012} (\bibinfo{year}{1997}).

\bibitem[{\citenamefont{Jelezko et~al.}(2004)\citenamefont{Jelezko, Gaebel,
  Popa, Gruber, and Wrachtrup}}]{Jelezko2004a}
\bibinfo{author}{\bibfnamefont{F.}~\bibnamefont{Jelezko}},
  \bibinfo{author}{\bibfnamefont{T.}~\bibnamefont{Gaebel}},
  \bibinfo{author}{\bibfnamefont{I.}~\bibnamefont{Popa}},
  \bibinfo{author}{\bibfnamefont{A.}~\bibnamefont{Gruber}}, \bibnamefont{and}
  \bibinfo{author}{\bibfnamefont{J.}~\bibnamefont{Wrachtrup}},
  \bibinfo{journal}{Physical Review Letters} \textbf{\bibinfo{volume}{92}},
  \bibinfo{pages}{076401} (\bibinfo{year}{2004}).

\bibitem[{\citenamefont{Balasubramanian
  et~al.}(2009)\citenamefont{Balasubramanian, Neumann, Twitchen, Markham,
  Kolesov, Mizuochi, Isoya, Achard, Beck, Tissler
  et~al.}}]{Balasubramanian2009}
\bibinfo{author}{\bibfnamefont{G.}~\bibnamefont{Balasubramanian}},
  \bibinfo{author}{\bibfnamefont{P.}~\bibnamefont{Neumann}},
  \bibinfo{author}{\bibfnamefont{D.}~\bibnamefont{Twitchen}},
  \bibinfo{author}{\bibfnamefont{M.}~\bibnamefont{Markham}},
  \bibinfo{author}{\bibfnamefont{R.}~\bibnamefont{Kolesov}},
  \bibinfo{author}{\bibfnamefont{N.}~\bibnamefont{Mizuochi}},
  \bibinfo{author}{\bibfnamefont{J.}~\bibnamefont{Isoya}},
  \bibinfo{author}{\bibfnamefont{J.}~\bibnamefont{Achard}},
  \bibinfo{author}{\bibfnamefont{J.}~\bibnamefont{Beck}},
  \bibinfo{author}{\bibfnamefont{J.}~\bibnamefont{Tissler}},
  \bibnamefont{et~al.}, \bibinfo{journal}{Nat Mater}
  \textbf{\bibinfo{volume}{8}}, \bibinfo{pages}{383} (\bibinfo{year}{2009}).

\bibitem[{\citenamefont{Taylor et~al.}(2008)\citenamefont{Taylor, Cappellaro,
  Childress, Jiang, Budker, Hemmer, Yacoby, Walsworth, and Lukin}}]{Taylor2008}
\bibinfo{author}{\bibfnamefont{J.~M.} \bibnamefont{Taylor}},
  \bibinfo{author}{\bibfnamefont{P.}~\bibnamefont{Cappellaro}},
  \bibinfo{author}{\bibfnamefont{L.}~\bibnamefont{Childress}},
  \bibinfo{author}{\bibfnamefont{L.}~\bibnamefont{Jiang}},
  \bibinfo{author}{\bibfnamefont{D.}~\bibnamefont{Budker}},
  \bibinfo{author}{\bibfnamefont{P.~R.} \bibnamefont{Hemmer}},
  \bibinfo{author}{\bibfnamefont{A.}~\bibnamefont{Yacoby}},
  \bibinfo{author}{\bibfnamefont{R.}~\bibnamefont{Walsworth}},
  \bibnamefont{and} \bibinfo{author}{\bibfnamefont{M.~D.} \bibnamefont{Lukin}},
  \bibinfo{journal}{Nat. Phys.} \textbf{\bibinfo{volume}{4}},
  \bibinfo{pages}{810} (\bibinfo{year}{2008}).

\bibitem[{\citenamefont{Maze et~al.}(2008)\citenamefont{Maze, Stanwix, Hodges,
  Hong, Taylor, Cappellaro, Jiang, Dutt, Togan, Zibrov et~al.}}]{Maze2008}
\bibinfo{author}{\bibfnamefont{J.~R.} \bibnamefont{Maze}},
  \bibinfo{author}{\bibfnamefont{P.~L.} \bibnamefont{Stanwix}},
  \bibinfo{author}{\bibfnamefont{J.~S.} \bibnamefont{Hodges}},
  \bibinfo{author}{\bibfnamefont{S.}~\bibnamefont{Hong}},
  \bibinfo{author}{\bibfnamefont{J.~M.} \bibnamefont{Taylor}},
  \bibinfo{author}{\bibfnamefont{P.}~\bibnamefont{Cappellaro}},
  \bibinfo{author}{\bibfnamefont{L.}~\bibnamefont{Jiang}},
  \bibinfo{author}{\bibfnamefont{M.~V.~G.} \bibnamefont{Dutt}},
  \bibinfo{author}{\bibfnamefont{E.}~\bibnamefont{Togan}},
  \bibinfo{author}{\bibfnamefont{A.~S.} \bibnamefont{Zibrov}},
  \bibnamefont{et~al.}, \bibinfo{journal}{Nature}
  \textbf{\bibinfo{volume}{455}}, \bibinfo{pages}{644} (\bibinfo{year}{2008}).

\bibitem[{\citenamefont{Waldherr et~al.}(2012)\citenamefont{Waldherr, Beck,
  Neumann, Said, Nitsche, Markham, Twitchen, Twamley, Jelezko, and
  Wrachtrup}}]{Waldherr2012}
\bibinfo{author}{\bibfnamefont{G.}~\bibnamefont{Waldherr}},
  \bibinfo{author}{\bibfnamefont{J.}~\bibnamefont{Beck}},
  \bibinfo{author}{\bibfnamefont{P.}~\bibnamefont{Neumann}},
  \bibinfo{author}{\bibfnamefont{R.~S.} \bibnamefont{Said}},
  \bibinfo{author}{\bibfnamefont{M.}~\bibnamefont{Nitsche}},
  \bibinfo{author}{\bibfnamefont{M.~L.} \bibnamefont{Markham}},
  \bibinfo{author}{\bibfnamefont{D.~J.} \bibnamefont{Twitchen}},
  \bibinfo{author}{\bibfnamefont{J.}~\bibnamefont{Twamley}},
  \bibinfo{author}{\bibfnamefont{F.}~\bibnamefont{Jelezko}}, \bibnamefont{and}
  \bibinfo{author}{\bibfnamefont{J.}~\bibnamefont{Wrachtrup}},
  \bibinfo{journal}{Nat. Nano.} \textbf{\bibinfo{volume}{7}},
  \bibinfo{pages}{105} (\bibinfo{year}{2012}).

\bibitem[{\citenamefont{H{\"a}berle et~al.}(2013)\citenamefont{H{\"a}berle,
  Schmid-Lorch, Karrai, Reinhard, and Wrachtrup}}]{Reinhard13}
\bibinfo{author}{\bibfnamefont{T.}~\bibnamefont{H{\"a}berle}},
  \bibinfo{author}{\bibfnamefont{D.}~\bibnamefont{Schmid-Lorch}},
  \bibinfo{author}{\bibfnamefont{K.}~\bibnamefont{Karrai}},
  \bibinfo{author}{\bibfnamefont{F.}~\bibnamefont{Reinhard}}, \bibnamefont{and}
  \bibinfo{author}{\bibfnamefont{J.}~\bibnamefont{Wrachtrup}},
  \bibinfo{journal}{Phys. Rev. Lett.} \textbf{\bibinfo{volume}{111}},
  \bibinfo{pages}{170801} (\bibinfo{year}{2013}).

\bibitem[{\citenamefont{Neumann et~al.}(2008)\citenamefont{Neumann, Mizuochi,
  Rempp, Hemmer, Watanabe, Yamasaki, Jacques, Gaebel, Jelezko, and
  Wrachtrup}}]{Neumann2008}
\bibinfo{author}{\bibfnamefont{P.}~\bibnamefont{Neumann}},
  \bibinfo{author}{\bibfnamefont{N.}~\bibnamefont{Mizuochi}},
  \bibinfo{author}{\bibfnamefont{F.}~\bibnamefont{Rempp}},
  \bibinfo{author}{\bibfnamefont{P.}~\bibnamefont{Hemmer}},
  \bibinfo{author}{\bibfnamefont{H.}~\bibnamefont{Watanabe}},
  \bibinfo{author}{\bibfnamefont{S.}~\bibnamefont{Yamasaki}},
  \bibinfo{author}{\bibfnamefont{V.}~\bibnamefont{Jacques}},
  \bibinfo{author}{\bibfnamefont{T.}~\bibnamefont{Gaebel}},
  \bibinfo{author}{\bibfnamefont{F.}~\bibnamefont{Jelezko}}, \bibnamefont{and}
  \bibinfo{author}{\bibfnamefont{J.}~\bibnamefont{Wrachtrup}},
  \bibinfo{journal}{Science} \textbf{\bibinfo{volume}{320}},
  \bibinfo{pages}{1326} (\bibinfo{year}{2008}).

\bibitem[{\citenamefont{Neumann}(2010)}]{Neumann2010}
\bibinfo{author}{\bibfnamefont{P.}~\bibnamefont{Neumann}},
  \bibinfo{journal}{Nat. Phys.} \textbf{\bibinfo{volume}{6}},
  \bibinfo{pages}{249} (\bibinfo{year}{2010}).

\bibitem[{\citenamefont{Fuchs et~al.}(2009)\citenamefont{Fuchs, Dobrovitski,
  Toyli, Heremans, and Awschalom}}]{Fuchs2009}
\bibinfo{author}{\bibfnamefont{G.~D.} \bibnamefont{Fuchs}},
  \bibinfo{author}{\bibfnamefont{V.~V.} \bibnamefont{Dobrovitski}},
  \bibinfo{author}{\bibfnamefont{D.~M.} \bibnamefont{Toyli}},
  \bibinfo{author}{\bibfnamefont{F.~J.} \bibnamefont{Heremans}},
  \bibnamefont{and} \bibinfo{author}{\bibfnamefont{D.~D.}
  \bibnamefont{Awschalom}}, \bibinfo{journal}{Science}
  \textbf{\bibinfo{volume}{326}}, \bibinfo{pages}{1520} (\bibinfo{year}{2009}).

\bibitem[{\citenamefont{Wrachtrup and Jelezko}(2006)}]{Wrachtrup2006b}
\bibinfo{author}{\bibfnamefont{J.}~\bibnamefont{Wrachtrup}} \bibnamefont{and}
  \bibinfo{author}{\bibfnamefont{F.}~\bibnamefont{Jelezko}},
  \bibinfo{journal}{Phys. Stat. Sol. A} \textbf{\bibinfo{volume}{203}},
  \bibinfo{pages}{3207} (\bibinfo{year}{2006}).

\bibitem[{\citenamefont{Nielsen and Chuang}(2000)}]{Nielsen2000}
\bibinfo{author}{\bibfnamefont{M.~A.} \bibnamefont{Nielsen}} \bibnamefont{and}
  \bibinfo{author}{\bibfnamefont{I.~L.} \bibnamefont{Chuang}},
  \emph{\bibinfo{title}{Quantum computation and quantum information}}
  (\bibinfo{publisher}{Cambridge University Press}, \bibinfo{year}{2000}).

\bibitem[{\citenamefont{Mehring and Webberu{\ss}}(2001)}]{MehringQC}
\bibinfo{author}{\bibfnamefont{M.}~\bibnamefont{Mehring}} \bibnamefont{and}
  \bibinfo{author}{\bibfnamefont{V.~A.} \bibnamefont{Webberu{\ss}}},
  \emph{\bibinfo{title}{Object-Oriented Magnetic Resonance}}
  (\bibinfo{publisher}{Academic Press}, \bibinfo{address}{London},
  \bibinfo{year}{2001}).

\bibitem[{\citenamefont{Boscain and Mason}(2006)}]{Boscain2006}
\bibinfo{author}{\bibfnamefont{U.}~\bibnamefont{Boscain}} \bibnamefont{and}
  \bibinfo{author}{\bibfnamefont{P.}~\bibnamefont{Mason}},
  \bibinfo{journal}{Journal of Mathematical Physics}
  \textbf{\bibinfo{volume}{47}}, \bibinfo{pages}{062101}
  (\bibinfo{year}{2006}).

\bibitem[{\citenamefont{Fortunato et~al.}(2002)\citenamefont{Fortunato, Pravia,
  Boulant, Teklemariam, Havel, and Cory}}]{Fortunato02}
\bibinfo{author}{\bibfnamefont{E.~M.} \bibnamefont{Fortunato}},
  \bibinfo{author}{\bibfnamefont{M.~A.} \bibnamefont{Pravia}},
  \bibinfo{author}{\bibfnamefont{N.}~\bibnamefont{Boulant}},
  \bibinfo{author}{\bibfnamefont{G.}~\bibnamefont{Teklemariam}},
  \bibinfo{author}{\bibfnamefont{T.~F.} \bibnamefont{Havel}}, \bibnamefont{and}
  \bibinfo{author}{\bibfnamefont{D.~G.} \bibnamefont{Cory}},
  \bibinfo{journal}{J. Chem. Phys.} \textbf{\bibinfo{volume}{116}},
  \bibinfo{pages}{7599} (\bibinfo{year}{2002}).

\bibitem[{\citenamefont{Slichter}(1996)}]{Slichter96}
\bibinfo{author}{\bibfnamefont{C.}~\bibnamefont{Slichter}},
  \emph{\bibinfo{title}{Principles of Magnetic Resonance}}
  (\bibinfo{publisher}{Springer-Verlag}, \bibinfo{year}{1996}).

\bibitem[{\citenamefont{Schweiger and Jeschke}(2001)}]{Schweiger2001}
\bibinfo{author}{\bibfnamefont{A.}~\bibnamefont{Schweiger}} \bibnamefont{and}
  \bibinfo{author}{\bibfnamefont{G.}~\bibnamefont{Jeschke}},
  \emph{\bibinfo{title}{Principles of pulse electron paramagnetic resonance}}
  (\bibinfo{publisher}{Oxford University Press}, \bibinfo{year}{2001}).

\bibitem[{\citenamefont{Maze et~al.}(2012)\citenamefont{Maze, Dreau,
  Waselowski, Duarte, Roch, and Jacques}}]{Maze12}
\bibinfo{author}{\bibfnamefont{J.~R.} \bibnamefont{Maze}},
  \bibinfo{author}{\bibfnamefont{A.}~\bibnamefont{Dreau}},
  \bibinfo{author}{\bibfnamefont{V.}~\bibnamefont{Waselowski}},
  \bibinfo{author}{\bibfnamefont{H.}~\bibnamefont{Duarte}},
  \bibinfo{author}{\bibfnamefont{J.-F.} \bibnamefont{Roch}}, \bibnamefont{and}
  \bibinfo{author}{\bibfnamefont{V.}~\bibnamefont{Jacques}},
  \bibinfo{journal}{New Journal of Physics} \textbf{\bibinfo{volume}{14}},
  \bibinfo{pages}{103041} (\bibinfo{year}{2012}).

\bibitem[{\citenamefont{Niemeyer et~al.}(2013)\citenamefont{Niemeyer, Shim,
  Zhang, Suter, Taniguchi, Teraji, Abe, Onoda, Yamamoto, Ohshima
  et~al.}}]{Suter13}
\bibinfo{author}{\bibfnamefont{I.}~\bibnamefont{Niemeyer}},
  \bibinfo{author}{\bibfnamefont{J.~H.} \bibnamefont{Shim}},
  \bibinfo{author}{\bibfnamefont{J.}~\bibnamefont{Zhang}},
  \bibinfo{author}{\bibfnamefont{D.}~\bibnamefont{Suter}},
  \bibinfo{author}{\bibfnamefont{T.}~\bibnamefont{Taniguchi}},
  \bibinfo{author}{\bibfnamefont{T.}~\bibnamefont{Teraji}},
  \bibinfo{author}{\bibfnamefont{H.}~\bibnamefont{Abe}},
  \bibinfo{author}{\bibfnamefont{S.}~\bibnamefont{Onoda}},
  \bibinfo{author}{\bibfnamefont{T.}~\bibnamefont{Yamamoto}},
  \bibinfo{author}{\bibfnamefont{T.}~\bibnamefont{Ohshima}},
  \bibnamefont{et~al.}, \bibinfo{journal}{New Journal of Physics}
  \textbf{\bibinfo{volume}{15}}, \bibinfo{pages}{033027}
  (\bibinfo{year}{2013}).

\bibitem[{\citenamefont{M. et~al.}(2012)\citenamefont{M., Momeen, and
  Dutt}}]{Nusran2012}
\bibinfo{author}{\bibfnamefont{N.~N.} \bibnamefont{M.}},
  \bibinfo{author}{\bibfnamefont{M.~U.} \bibnamefont{Momeen}},
  \bibnamefont{and} \bibinfo{author}{\bibfnamefont{M.~V.~G.}
  \bibnamefont{Dutt}}, \bibinfo{journal}{Nat. Nano.}
  \textbf{\bibinfo{volume}{7}}, \bibinfo{pages}{109} (\bibinfo{year}{2012}).

\bibitem[{\citenamefont{Fuchs et~al.}(2012)\citenamefont{Fuchs, Falk,
  Dobrovitski, and Awschalom}}]{Fuchs12}
\bibinfo{author}{\bibfnamefont{G.~D.} \bibnamefont{Fuchs}},
  \bibinfo{author}{\bibfnamefont{A.~L.} \bibnamefont{Falk}},
  \bibinfo{author}{\bibfnamefont{V.~V.} \bibnamefont{Dobrovitski}},
  \bibnamefont{and} \bibinfo{author}{\bibfnamefont{D.~D.}
  \bibnamefont{Awschalom}}, \bibinfo{journal}{Phys. Rev. Lett.}
  \textbf{\bibinfo{volume}{108}}, \bibinfo{pages}{157602}
  (\bibinfo{year}{2012}).

\bibitem[{\citenamefont{de~Lange et~al.}(2012)\citenamefont{de~Lange, van~der
  Sar, Blok, Wang, Dobrovitski, and Hanson}}]{Hanson12}
\bibinfo{author}{\bibfnamefont{G.}~\bibnamefont{de~Lange}},
  \bibinfo{author}{\bibfnamefont{T.}~\bibnamefont{van~der Sar}},
  \bibinfo{author}{\bibfnamefont{M.}~\bibnamefont{Blok}},
  \bibinfo{author}{\bibfnamefont{Z.-H.} \bibnamefont{Wang}},
  \bibinfo{author}{\bibfnamefont{V.}~\bibnamefont{Dobrovitski}},
  \bibnamefont{and} \bibinfo{author}{\bibfnamefont{R.}~\bibnamefont{Hanson}},
  \bibinfo{journal}{SCIENTIFIC REPORTS} \textbf{\bibinfo{volume}{2}},
  \bibinfo{pages}{382} (\bibinfo{year}{2012}).

\bibitem[{\citenamefont{Hanh}(1950)}]{Hahn50}
\bibinfo{author}{\bibfnamefont{E.~L.} \bibnamefont{Hanh}},
  \bibinfo{journal}{Phys. Rev.} \textbf{\bibinfo{volume}{80}},
  \bibinfo{pages}{580} (\bibinfo{year}{1950}).

\bibitem[{\citenamefont{Forrer et~al.}(2004)\citenamefont{Forrer, Schmutz,
  Tschaggelar, and Schweiger}}]{Schweiger04}
\bibinfo{author}{\bibfnamefont{J.}~\bibnamefont{Forrer}},
  \bibinfo{author}{\bibfnamefont{H.}~\bibnamefont{Schmutz}},
  \bibinfo{author}{\bibfnamefont{R.}~\bibnamefont{Tschaggelar}},
  \bibnamefont{and}
  \bibinfo{author}{\bibfnamefont{A.}~\bibnamefont{Schweiger}},
  \bibinfo{journal}{J. Magn. Reson.} \textbf{\bibinfo{volume}{166}},
  \bibinfo{pages}{246} (\bibinfo{year}{2004}).

\bibitem[{\citenamefont{Tseitlin et~al.}(2011)\citenamefont{Tseitlin, Quine,
  Rinard, Eaton, and Eaton}}]{Eaton11}
\bibinfo{author}{\bibfnamefont{M.}~\bibnamefont{Tseitlin}},
  \bibinfo{author}{\bibfnamefont{R.~W.} \bibnamefont{Quine}},
  \bibinfo{author}{\bibfnamefont{G.~A.} \bibnamefont{Rinard}},
  \bibinfo{author}{\bibfnamefont{S.~S.} \bibnamefont{Eaton}}, \bibnamefont{and}
  \bibinfo{author}{\bibfnamefont{G.}~\bibnamefont{Eaton}}, \bibinfo{journal}{J.
  Magn. Reson.} \textbf{\bibinfo{volume}{213}}, \bibinfo{pages}{119}
  (\bibinfo{year}{2011}).

\end{thebibliography}

\begin{thebibliography}{1}

\bibitem{Boscain2006}
Ugo Boscain and Paolo Mason.
\newblock Time minimal trajectories for a spin 1/2 particle in a magnetic
  field.
\newblock {\em Journal of Mathematical Physics}, 47(6):062101, 2006.

\bibitem{Caneva2011}
Tommaso Caneva, Tommaso Calarco, and Simone Montangero.
\newblock Chopped random-basis quantum optimization.
\newblock {\em Phys. Rev. A}, 84:022326, 2011.

\bibitem{Doria2011}
Patrick Doria, Tommaso Calarco, and Simone Montangero.
\newblock Optimal control technique for many-body quantum dynamics.
\newblock {\em Phys. Rev. Lett.}, 106:190501, 2011.

\bibitem{granicher1996messung}
W.H. Gr{\"a}nicher.
\newblock {\em Messung beendet - was nun?: Einf{\"u}hrung und Nachschlagewerk
  f{\"u}r die Planung und Auswertung von Messungen}.
\newblock vdf, Hochschulverlag AG an der ETH Z{\"u}rich, 1996.

\bibitem{nielsenchuang}
Michael~A. Nielsen and Isaac~L. Chuang.
\newblock {\em {Quantum Computation and Quantum Information}}.
\newblock Cambridge University Press, 2010.

\end{thebibliography}

\end{document}